# Broadband Coherent Wave Control Through Photonic Collisions at Time Interfaces

Emanuele Galiffi[1]†, Gengyu Xu[1]†, Shixiong Yin[1,2], Hady Moussa[1,2], Younes Ra'di[1,4] and Andrea Alù[1,2,3]*

[1]Photonics Initiative, Advanced Science Research Center, City University of New York, 85 St. Nicholas Terrace, 10031, New York, United States

[2]Department of Electrical Engineering, City College, City University of New York

[3]Physics Program, Graduate Center, City University of New York

[4]Department of Electrical Engineering and Computer Science, Syracuse University, Syracuse, NY 13244, United States

*Corresponding author. Email: aalu@gc.cuny.edu

† These authors contributed equally to the work.

**Abstract:** Coherent wave control exploits the interference among multiple waves impinging on a system to suppress or enhance outgoing signals based on their relative phase and amplitude. This process inherently requires the combination of non-Hermiticity, in order to enable energy exchanges among the waves, and spatial interfaces in order to tailor their scattering. Here we explore the temporal analogue of this phenomenon, based on time-interfaces that support instantaneous non-Hermitian scattering events for photons. Based on this mechanism, we demonstrate ultra-broadband temporal coherent wave control and the photonic analogue of mechanical collisions with phase-tunable elastic features. We apply them to erase, enhance and reshape arbitrary pulses by suitably tailoring the amplitude and phase of counterpropagating signals. Our findings provide a pathway to effectively sculpt broadband light with light without requiring spatial boundaries, within an ultrafast and low-energy platform.

**One-Sentence Summary:** Temporal coherent wave control and collisions between photons are introduced and demonstrated at time-interfaces in a transmission-line metamaterial.

**Main Text:**

When propagating through a lossless medium, counterpropagating photons do not interact, but merely pass through each other. However, when scattering off a lossy structure, the orthogonality between them is broken, and energy exchanges can occur, so that they may be entirely transmitted or absorbed based solely on their relative amplitude and phase. Such coherent wave control, which inherently requires non-Hermitian scattering, has been at the basis of exciting developments in the context of wave physics. A landmark example is coherent perfect absorption, i.e., the destructive interference of all outgoing waves due to their energy being entirely dissipated in a non-Hermitian system – a process analogous to time-reversed lasing (1,2), as depicted in the left panel of Fig. 1A.

Here we introduce and experimentally demonstrate coherent wave control (CWC) at temporal interfaces: we induce non-Hermiticity in a lossless photonic system by abruptly changing its electromagnetic properties uniformly in space, realizing a time-interface (TI) capable of inducing extremely fast energy exchanges between two counterpropagating waves. In temporal CWC (Fig. 1A, right), the time-reflected waves produced by a TI can interfere destructively with the opposite time-refracted waves, enabling the cancellation of one (shown in Fig. 1A and in our experiments) or even both outgoing waves. Remarkably, the underlying platform can be fully Hermitian before and after the temporal scattering event, whereas energy exchanges occur only for the ultrashort duration of the TI, allowing us to instantaneously control over very broad bandwidths the temporal evolution of the incoming signals by suitably tailoring amplitude and phase of a counterpropagating idler pulse. The resulting photon-photon interactions bear a correspondence with mechanical collisions, whereby the total momentum in the system is conserved, while the degree of inelasticity can be tuned from destructive to constructive via the relative phase at the TI, which can act as an internal non-conservative force between them.

Following a seminal demonstration for water waves (3), TIs have only recently found their implementation for electromagnetic waves in a transmission-line metamaterial (TLM) (4), whose effective permittivity can be uniformly modified as strongly as 100% within less than 3 ns by densely packed arrays of shunt capacitors. In this platform, TIs give rise to efficient temporal scattering characterized by causal relations fundamentally different from spatial scattering phenomena (5,5), establishing the building blocks to realize time-metamaterials (7,8) and photonic time-crystals (9). In turn, these exotic systems host enticing forms of wave-matter interaction, including nonreciprocal wave propagation (10-12), amplification (13,14) and drag (15), time-refraction (16, 17), time-diffraction (18, 19), temporal localization (20-22), Hermitian (23-25) and non-Hermitian (26) Floquet topological physics, particularly exploiting synthetic frequency dimensions (27), temporal parity-time symmetry (28), temporal wavenumber filtering (29) and pulse aiming via anisotropic TIs (30), as well as imaging through disordered media (31), negative refraction (32,33) and subdiffractional focusing (34), among several others (6). Here, we use a similar platform to demonstrate ultrabroadband CWC and ultrafast pulse shaping at TIs.

Consider two counterpropagating monochromatic electromagnetic waves of identical frequency $\omega_1$ and wavenumber $k$ propagating in a homogeneous medium. At time instant $t = 0$, the wave impedance is uniformly switched from $Z_1$ to $Z_2$ (Fig. 1B, left), where we assume $Z_2 < Z_1$ without loss of generality (see SM Sec. 1 for details). The right panel of Fig. 1B shows the corresponding scattering process: if this abrupt temporal inhomogeneity is significantly faster than the temporal oscillation period of the involved signals (35), the resulting TI generates, from each incoming signal (dotted lines), a time-refracted (continuous) and a time-reflected (dashed) one at a new frequency but identical wavelength. The respective displacement-field scattering coefficients are $T = (Z_2 + Z_1)/2Z_2 > 0$ and $R = (Z_2 - Z_1)/2Z_2 < 0$ (4, 5), and the phase accumulation is negligible. The total amplitudes of the waves after the TI then read $a_2^+ = Ta_1^+ + Ra_1^- e^{i\phi}$ and $a_2^- = Ta_1^- e^{i\phi} + Ra_1^+$ respectively, where $a_1^+, a_1^-$ denote the magnitudes of the displacement-field of the forward (+) and backward (-) incoming waves, and $\phi$ is the relative phase between them at the TI.

The effect of the instantaneous non-Hermiticity introduced by the TI can be tuned via the relative phase $\phi$. Figure 1 C-H show three possible scenarios corresponding to a phase (C, D) $\phi = 0$, (E, F) $\pi/2$, or (G, H) $\pi$ between the two incoming waves. When they are in phase at the TI (C), the phase flip upon time-reflection (TR) implies that the two outgoing waves in both the positive and negative direction will be out of phase ($R < 0$), so that the interaction will be destructive, minimizing the total energy in the system, while preserving the total electromagnetic momentum density $S = Z_1(|a_1^+|^2 - |a_1^-|^2)$, which must be conserved due to the translational invariance of the system. As a special scenario in this regime, if the two amplitudes obey $a_1^- = (T/|R|)a_1^+$, the outgoing backward wave is completely erased (Fig. 1D). Energetically, this process results in the minimum total energy after the TI that ensures preservation of the momentum in the system. It is also possible to cancel both forward and backward waves, in a phenomenon analogous to CPA, by sending two waves of equal amplitude, but also requiring large impedance contrast $Z_2 \ll Z_1$ or $Z_2 \gg Z_1$, in which limit $|R|/|T| \to 1/2$.

Different choices of $\phi$ yield different total energy in the outgoing waves. If (E, F) $\phi = \pi/2$, the interaction is elastic, resulting from a balance between constructive and destructive interactions, which yields the same final energy $\bar{E}_2^{(tot)}$ as before the interaction (see SM Sec. S2 for details). Finally, perfectly constructive interactions occur (G, H) if the two waves are out of phase with each other at the TI ($\phi = \pi$), as phase-flipped TR-waves constructively interfere with the time-refracted ones in each direction. These different scattering regimes are shown as a function of the phase $\phi$ in Fig. 1I, exhibiting the characteristic signature of CWC: green lines indicate the final energy density in the forward (dashed), backward (dotted), and combined (continuous) waves upon TI-induced scattering in the presence of both interfering waves, whereas black lines indicate the corresponding values when the two waves do not interfere in the absence of TI. Constructive photonic collision regimes (pink background, $|\phi| > \pi/2$) are separated from destructive ones (beige background $|\phi| < \pi/2$) by the elasticity condition $|\phi| = \pi/2$ (light blue). Thus, the amplitude and phase of a counterpropagating idler signal can be tailored to perform the desired CWC operation on a forward propagating signal. Here CWC does not require any spatial interface or resonant buildup and, thanks to the infinitesimal duration of the TI, no phase accumulates during the temporal scattering process. Hence, temporal CWC is frequency agnostic, inherently ultrabroadband, somewhat analogous to the use of ultrathin metasurfaces for spatial CWC (36).

An interesting analogy also emerges with the physics of particle collisions: two colliding bodies that stick together upon impact due to the action of internal forces minimize the total final energy in the system (shown pictorially in Fig. 1J). By contrast, after an elastic collision (Fig. 1K) they move away from each other with equal center-of-mass velocities, conserving kinetic energy. Finally (Fig. 1L) if the internal force conveys additional energy to them, e.g., by releasing a pre-loaded massless spring between them, they will scatter away with increased total kinetic energy. Remarkably, the internal energy added to (or subtracted from) the two electromagnetic waves is provided here by the TI itself, which opens an infinitesimal window of time over which non-Hermiticity is introduced in the system, enabling the interaction between two otherwise orthogonal waves.

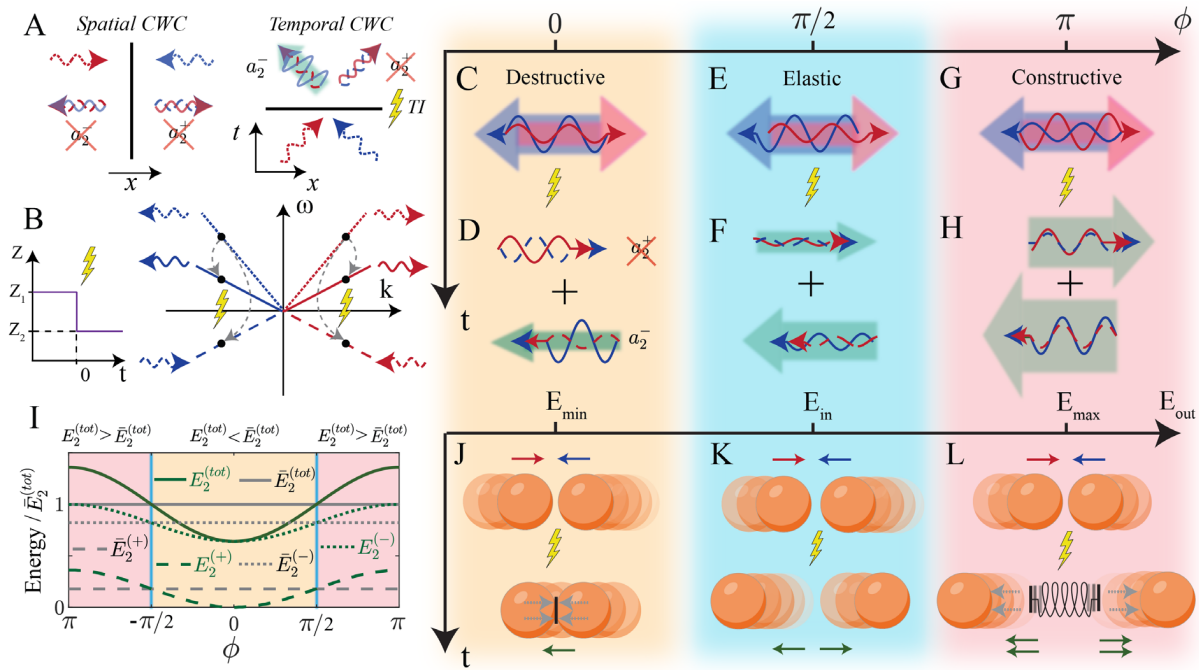

**Fig. 1:** Temporal coherent wave control and photonic collisions enabled by time-interfaces. **(A)** Spatial CWC (left) engages two waves impinging from opposite sides of a non-Hermitian scatterer, leading to their complete absorption. By contrast, temporal CWC (right) results from two counterpropagating waves undergoing partial time-reflection at a TI, enabling cancellation of one (shown experimentally in this work) or in principle both outgoing waves. In this case, the medium is Hermitian before and after the scattering, its non-Hermiticity residing in the infinitesimal TI itself. **(B)** Illustration of impedance switching at a TI (left), and the corresponding phase-space diagram (right), showing a spectral representation of temporal CWC, with incoming (dotted) forward (red) and backward (blue) waves being respectively scattered onto time-refracted (continuous) and time-reflected (dashed) waves. **(C-H)** A relative phase **(C-D)** $\phi = 0$, **(E-F)** $\pi/2$ or **(G-H)** $\pi$ between the two counterpropagating waves at the TI results in destructive (energy is reduced, potentially suppressed entirely), elastic (energy is conserved) or constructive (energy is maximized) photon-photon interactions respectively. Net momentum is conserved in all scenarios. **(I)** Final energy in forward ( $E_2^{(+)} = |a_2^+|^2$, green dashed), backward ( $E_2^{(-)} = |a_2^-|^2$, green dotted), and combined ( $E_2^{(tot)} = (|a_2^+|^2 + |a_2^-|^2)$ ) waves as a function of the relative phase at the TI, normalized to the total final energy $\bar{E}_2^{(tot)}$ in the no-interaction case, and compared to the energy in the respective combined forward ( $\bar{E}_2^{(+)}$ ) and backward ( $\bar{E}_2^{(-)}$ ) propagating waves without interaction (gray). The beige-shaded area denotes the phase regime $|\phi| < \pi/2$ over which the interaction is destructive, separated from the constructive regime $|\phi| > \pi/2$ (pink shading) by the elastic condition $|\phi| = \pi/2$ (light blue). Here we use $Z_1 = 1$, $Z_2 = 0.5$, $a_1^+ = 1$ and $|a_1^-| = 3$ with normalized units. **(J-L)** Centre-of-mass-frame illustration of photonic collisions corresponding to the three TI phases. The energy minimization **(J)** occurring when $\phi = 0$ is analogous to an inelastic collision, whereby the two bodies are stuck with each other via internal forces, minimizing the total energy. An elastic collision **(K)** results in the two bodies bouncing off each other exchanging energy and momentum but leaving both unchanged overall. Finally, if pushed apart by an internal force such as a pre-loaded spring, both colliding bodies scatter away with increased energy **(L)**, while preserving the total momentum in the system.

To observe photon collisions and demonstrate temporal CWC in a practical setup, we designed a transmission-line metamaterial (TLM) consisting of a 6.24-meter meandered microwave transmission line loaded with a subwavelength linear array of switches, with period of 0.208m, each connecting the transmission-line with a shunt lumped capacitor. The switches can be activated with a step-like bias voltage trigger. A simple schematic of our periodically loaded TLM, an upgraded version of our previous setup [4] allowing for two synchronized input ports, is shown in Fig. 2A: a broadband forward *signal* pulse (red) and a broadband backward *idler* pulse (blue) containing the same frequency spectrum are sent from the left and right ports of the TLM respectively, and measured at the $V_1$ and $V_2$ voltage probes as they enter and exit the system. When the two pulses overlap, the bias voltage triggers all switches synchronously and over time scales much faster (less than 4 ns) than the relevant frequencies of the incoming signals, so that the effective impedance and refractive index of the host medium are simultaneously and abruptly switched along the entire TLM. A photo and the specification details of our in-house fabricated device are provided in the SM Sec. S3.

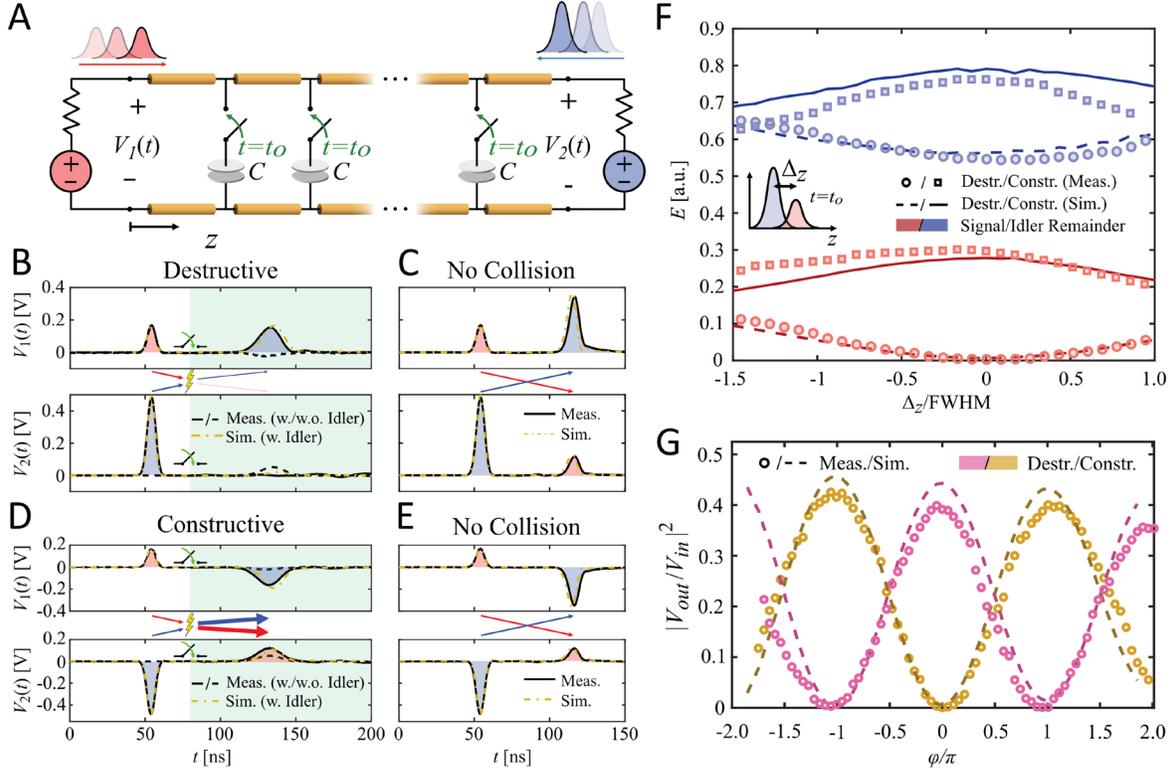

**Fig. 2.** Experimental implementation of broadband temporal CWC and photonic collisions in a TLM. **(A)** Schematic of the experimental platform. A TL periodically loaded by switched capacitors is excited from both ports with a right-propagating signal pulse (red) and a left-propagating idler pulse (blue). When the two pulses overlap, all switches are synchronously closed via a fast bias voltage, abruptly changing the effective impedance of the entire TLM, and inducing energy exchanges between the two pulses. **(B-E)** Experimental measurements: continuous and dashed black lines correspond to the experiment with and without idler pulse respectively, while dot-dashed yellow lines correspond to numerical simulations (with idler). **(B)** A destructive photonic collision, resulting in the complete annihilation of the signal pulse, as shown by the absence of a transmitted signal in the measured voltages $V_2$ at the right port. The remaining idler pulse measured by the $V_1$ probe, exhibits temporal broadening due to the frequency translation caused by the change in the refractive index of the TLM. Panel **(C)** shows the scenario with no collision (no switching), whereby the injected pulses travel through each other without interactions, and are transmitted to the opposite ports after some attenuation (see SM Sec. S3). **(D)** Switching the idler polarity is equivalent to shifting the phase of each Fourier component of the idler pulse by $\pi$, allowing us to induce constructive collisions, resulting in the final signal at the $V_2$ port showing a similar peak amplitude as the no-collision case **(E)**, yet with increased pulse duration, and doubled amplitude compared to the case without idler. **(F)** By varying the switching time, we can detune the spatial delay $\Delta_z$ between the two pulses upon collision, and hence continuously control the final energy in the signal (red) and idler (blue) pulses, manipulating the type and degree of inelasticity in the collision. Here markers and lines denote measurements and simulations, with circles (squares) and dashed (continuous) lines corresponding to input pulses with identical (opposite) polarity. **(G)** By considering a monochromatic signal with 3.21 rad/m carrier wave and spatial FWHM 1.86 m, we measure the periodic dependence of the final amplitudes with respect to the switching delay, reproducing Fig. 1I experimentally (markers) and numerically (dashed lines). Note how pulses with identical (yellow) and opposite (pink) polarity exhibit opposite phase dependence, demonstrating coherent control of the photonic collision.

Figures 2B-E show our measurements with (continuous black lines) and without (dashed black lines) idler, as well as numerical simulations (with idler, dot-dashed yellow lines, see SM Sec. 4 for details). In particular, Figure 2B refers to a maximally destructive photonic collision: the incoming broadband pulses (on the left) are sent in phase from the opposite ports, and the switches are activated when the pulses overlap in space, resulting in the complete annihilation of the forward-traveling signal (red) pulse upon the collision process. By contrast, if no switching occurs (Fig. 2C) the two pulses simply pass through each other without interacting. As a by-product, we observe how after the TI the pulse signal broadens in time due to the broadband frequency translation produced by the abrupt change in the effective refractive index of the TLM.

Conversely, Fig. 2D shows an instance of constructive inelastic collision: our broadband pulses are now sent in with opposite phase, by flipping the polarity of the idler signal. After the TI, the forward signal is amplified relative to the no-interference case (dashed line), demonstrating the opposite scenario, due to the change in relative phase for the same TI. Fig. 2E shows the corresponding no-collision scenario in the absence of a time interface. A discussion on measurement and data interpretation, as well as experimental results with high-frequency signal and idler pulses can be found in SM Sec. S5. Note that our specific implementation of a TI does not impart energy into the system (4), so all energy comparisons are relative to the case where the waves propagate in the switched medium independently of each other (see SM Sec. S2 for details).

To demonstrate coherent control of photon collisions at our time interface, we evaluate the integrated energy in the outgoing pulses as a function of the relative phase upon switching (see SM Sec. S6 for details). In Fig. 2F, we plot the measured (markers) and simulated (lines) energy in the forward (red) and backward (blue) pulses after the time interface against the spatial delay $\Delta_z$ between their two peaks at the TI, for pulses with equal (circles, dashed lines) and opposite (squares, continuous lines) polarity. Indeed, for signal and idler with the same phase we observe signal annihilation when the pulses overlap upon collision, and enhancement for opposite phases. Figure 2G shows the response for a single spatial Fourier component of the outgoing signal pulse, whose transmittance is plotted as a function of the relative phase $\phi$ between the corresponding Fourier amplitudes in the two pulses at the TI (see SM Sec. S7 for details). Pink and yellow data correspond to pulses with matching and inverted phase respectively, showing the periodic dependence predicted in Fig. 1I. These results demonstrate temporal coherent control of light in time metamaterials, with exciting implications that may be broadly extended in the context of non-Hermitian physics.

Finally, we propose an application of temporal CWC for distributed instantaneous pulse shaping, as demonstrated in Fig. 3. Given an incoming signal pulse, we can use a tailored opposite-traveling idler pulse to chisel away unwanted portions of the signal by means of a TI. This allows us to shape the entire pulse in a broadband and ultrafast platform, effectively sculpting light with light. Fig. 3A shows numerical simulations (SM Sec. S4) that demonstrate this concept: given (top two panels) a forward-traveling square signal pulse (red), we aim at erasing a specific portion (enclosed in the dashed box). Hence, we tailor a corresponding idler pulse (blue), shaped as the portion of the signal that we want to chisel away. As the two pulses overlap (middle panel), the TLM is switched, annihilating the undesired portion of the signal (bottom two panels). Figure 3B shows our experimental demonstration of this concept, reporting the measured voltages at the left and right probes. The incoming signal (red, top panel) is accurately trimmed by the idler traveling in the opposite direction (blue, bottom panel), as shown by the measured red-shaded curve in the bottom panel. The overlaid dashed line shows the shape of the final signal pulse in the absence of idler, highlighting the erased portions of the signal.

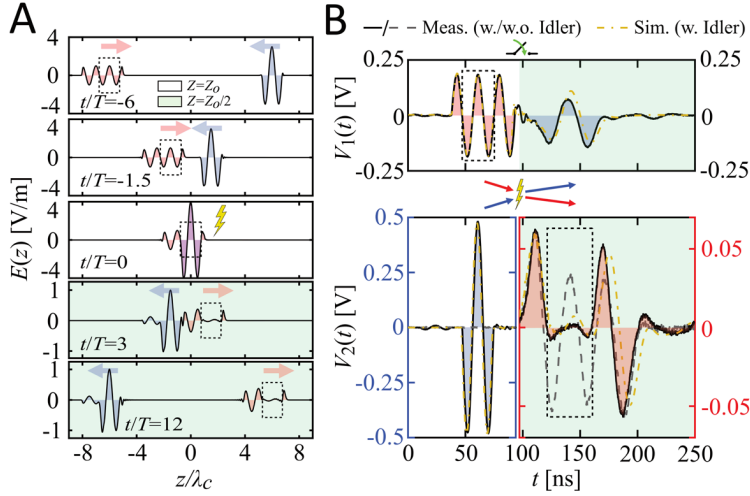

**Fig. 3.** Distributed instantaneous pulse shaping based on photonic collisions at a TI. **(A)** Simulations show an incoming signal pulse (red) sculpted by switching the TLM as it overlaps with an idler (blue) tailored to erase the portion of the signal delimited by the dashed box. Different panels refer to subsequent times. **(B)** Corresponding experimental measurements as a function of time at the two voltage probes. The undesired portion of the incoming signal is erased upon time-switching by the collision with the idler pulse, before the final sculpted signal travels to the right voltage probe $V_2$. The left and right axes in the bottom panel refer to idler and signal measurements, respectively.

In this work, we have introduced and experimentally demonstrated broadband temporal coherent wave control and photon-photon collisions enabled by frequency-agnostic temporal interfaces, whose nature can be tailored through the relative phase of the colliding waves. We have shown elastic, destructive and constructive collisions between photons, despite their neutral, bosonic nature. Based on this phenomenon, we can induce extreme energy exchanges between arbitrary pulses, realizing an ultrafast and ultrabroadband form of coherent wave control. Finally, we have shown an application of this concept to shape an incoming signal by colliding it with an idler tailored to erase or manipulate specific portions of it, effectively using light to sculpt light in an instantaneous and ultrabroadband fashion, opening opportunities for ultrafast pulse shaping. Our demonstration considers non-dispersive materials, but we envision enhanced opportunities arising in the case of dispersive systems, for which multiple output beams oscillating at different frequencies may be generated with tailored profiles (37).

While we have demonstrated this concept in an electromagnetic setting by implementing a transmission-line loaded metamaterial, this concept is general and can thus be readily translated to other classical or quantum wave settings. We expect that this result may open exciting opportunities for Floquet many-body systems (38), as well as non-Hermitian physics and extreme wave scattering phenomena in the presence of multiple input waves and of multiple time interfaces, as in the case of time crystals (9, 39), unveiling new forms of light-matter interactions, and applications for parallelized pulse shaping, energy mixing, conversion, and harvesting, as well as wave cancellation, not only in the microwave regime, but also at higher frequencies by leveraging high-speed graphene transistors (40), flash ionization in plasmas (41) or giant optical nonlinearities in epsilon-near-zero materials (42).

## Acknowledgments

This work was supported by the Air Force Office of Scientific Research, the CHARM program and the Simons Foundation. E.G. was supported by the Simons Foundation through a Junior Fellowship of the Simons Society of Fellows (855344, EG).

27. A. Dutt, Q. Lin, L. Yuan, M. Minkov, M. Xiao, S. Fan (2020). A single photonic cavity with two independent physical synthetic dimensions. *Science*, **367**(6473), 59-64.

28. H. Li, S. Yin, E. Galiffi, A. Alù (2021). Temporal parity-time symmetry for extreme energy transformations. *Physical Review Letters*, **127**(15), 153903.

29. V. Pacheco-Peña, N. Engheta (2020). Antireflection temporal coatings. *Optica*, **7**(4), 323-331.

30. V. Pacheco-Peña, N. Engheta (2020). Temporal aiming. *Light: Science & Applications*, **9**(1), 1-12.

31. A. P. Mosk, A. Lagendijk, G. Lerosey, M. Fink (2012). Controlling waves in space and time for imaging and focusing in complex media. *Nature photonics*, **6**(5), 283-292.

32. J. B. Pendry (2008). Time reversal and negative refraction. *Science*, **322**(5898), 71-73.

33. S. Vezzoli, V. Bruno, C. DeVault, T. Roger, V. M. Shalaev, A. Boltasseva, M. Ferrera, M. Clerici, A. Dubietis, D. Faccio (2018). Optical time reversal from time-dependent epsilon-near-zero media. *Physical Review Letters*, **120**(4), 043902 (2018).

34. G. Lerosey, J. De Rosny, A. Tourin, M. Fink (2007). Focusing beyond the diffraction limit with far-field time reversal. *Science*, **315**(5815), 1120-1122.

35. E. Galiffi, S. Yin A. Alù (2022) Tapered photonic switching. *Nanophotonics* **11**(16), 3575-3581.

36. T. Roger, S. Vezzoli, E. Bolduc, J. Valente, J. J. Heitz, J. Jeffers, C. Soci, J. Leach, C. Couteau, N. Zheludev, D. Faccio (2015). Coherent perfect absorption in deeply subwavelength films in the single-photon regime. *Nature communications*, **6**(1), 1-5.

37. D. M. Solís, R. Kastner, N. Engheta (2021). Time-varying materials in the presence of dispersion: plane-wave propagation in a Lorentzian medium with temporal discontinuity. *Photonics Research*, **9**(9), 1842-1853.

38. T. Kitagawa, E. Kitagawa, M. Rudner, E. Demler (2010), *Physical Review B*, **82**(23), 235114.

39. X. Mi, et al. (2022). Time-crystalline eigenstate order on a quantum processor. *Nature*, **601**(7894), 531-536.

40. M. Ono, M. Hata, M. Tsunekawa, K. Nozaki, H. Sumikura, H. Chiba, M. Notomi (2020). Ultrafast and energy-efficient all-optical switching with graphene-loaded deep-subwavelength plasmonic waveguides. *Nature Photonics*, **14**(1), 37-43.

41. A. Nishida, N. Yugami, T. Higashiguchi, T. Otsuka, F. Suzuki, M. Nakata, Y. Sentoku & R. Kodama (2012). Experimental observation of frequency up-conversion by flash ionization. *Applied Physics Letters*, **101**(16), 161118.

42. M. Z. Alam, I. De Leon, R. W. Boyd (2016). Large optical nonlinearity of indium tin oxide in its epsilon-near-zero region. *Science*, **352**(6287), 795-797.
11